%Paper: 9203045
%From: lopez@itp.ethz.ch (Martina Lopez)
%Date: Wed, 18 Mar 92 11:44:47 +0100

\magnification=\magstep1
\vsize=24truecm
\hsize=16truecm
\overfullrule=0pt
\baselineskip=.65truecm

\voffset .5truecm

\rightline{ETH-TH/92-6}
\vskip 2truecm

\centerline{{\bf RESULTS OF THE CLASSIFICATION OF SUPERCONFORMAL}}

\centerline{{\bf ALGEBRAS IN TWO DIMENSIONS}}

\vskip 0.5truecm

\centerline{\bf E.S. Fradkin\rm\footnote{$^*$}{present  address: Institut f\"ur
Theoretische Physik, ETH-H\"onggerberg, CH-8093 Z\"urich} and
\bf V.Ya. Linetsky\rm}

\vskip 0.5truecm

\centerline{Department of Theoretical Physics,}

\centerline{P.N.Lebedev Physical Institute, Leninsky pr. 53, Moscow 117924}

\vskip 3truecm

\centerline{{\bf Abstract}}

\vskip 0.5truecm

A list of superconformal chiral operator product expansion
algebras with quadratic nonlinearity in two dimensions is completed on the
basis of the known classification of little conformal Lie superalgebras. In
addition to the previously known cases and the constructed in our previous
paper exceptional $N=8$ superalgebra associated with $F(4)$, a novel
exceptional $N=7$ superconformal algebra associated with $G(3)$ is found, as
well as a whole family of superalgebras containing affine $\widehat{su}_2
\oplus
\widehat{usp}_{2N}$. A classification scheme for quasisuperconformal algebras
is
also outlined.
\vskip 3truecm
\centerline{February 1992}

\vfill\eject

Recently we have found a novel $N=8$ superconformal algebra in two dimensions
[1] with supercurrents in the eight-dimensional spinor representation of
Spin(7). It involves a peculiar composite :{\it JJ}: term in the anticommutator
of two supercurrents, similar to the superconformal algebras of Refs.[2-5]. It
has encouraged us to see whether there exist any other previously unknown
superconformal algebras involving quadratic composites (simple superconformal
algebras of the Lie type involving operators with scale dimensions 2, $3/2$, 1
and $1/2$ have been classified by Ramond and Schwarz [6]). As a result we have
completed a list of superconformal algebras of this category which is given in
the present letter. Here we present only the final results, all the proofs and
details being postponed to a more detailed paper [7].

Suppose we are given a pair $(g,\rho )$ of a compact Lie algebra $g$ and its
representation $\rho $. Consider a two-dimensional conformal field theory with
an energy-momentum $T(z)$, a multiplet of supercurrents $G^\alpha (z)$ in the
representation $\rho $ of $g$ and a multiplet of Kac-Moody currents $J^A(z)$ in
the adjoint representation of $g$, scale dimensions being 2, $3/2$ and 1,
respectively. (We are concerned with the holomorphic sector only.)

Then a problem we address is to enumerate all the admissible pairs $(g,\rho )$
such that the corresponding associative ${\bf Z}_2$-graded chiral operator
product expansion closes on $T$, $G$, $J$ and their (normal ordered) composites
only, and to evaluate explicitly the corresponding superconformal algebras. We
assume the resulted superconformal algebras, which we sometimes denote
$SC(g,\rho )$, to be simple in the following sense.
Consider an associative algebra $U$ with generating elements
$\{L_n,G^\alpha _r$, $J^A_n\}$ (the modes of $T(z)$, $G^\alpha (z)$ and
$J^A(z))$ and defining relations determined by the corresponding OPE. Then a
{\it SC} is said to be simple iff $U$ contains no non-trivial ideals generated
by a non-trivial set of elements linear in generating elements. In case when
{\it SC} is of the Lie-type $(i.e$. involvs no composites in the
supercommutators) this definition is equivalent to the usual definition of
simplicity of a Lie superalgebra and $U$ is its universal enveloping.

Actually, we have proved that there exists a simple $SC(g,\rho )$ iff there
exists a simple little (finite-dimensional) conformal Lie superalgebra
$LSC(g,\rho )$ with its even subalgebra isomorphic to $s\ell (2;{\bf R})
\oplus  g$ and its odd subspace transforming in the representation $(2,\rho )$
of $s\ell (2;{\bf R}) \oplus  g$. This proposition is quite non-trivial since
the {\it LSC} is not generally, $i.e$. when composites are presented, a
subalgebra of {\it SC} (the situation here is similar to, $e.g.$, the relation
of $s\ell _N$ to $W_N)$ and requires a proof based on the detailed analysis of
the Jacobi identities for the quantum algebra {\it SC}. (For how to extract
{\it LSC} from {\it SC} in non-linear case see Ref.[1].) However, we will not
discuss it here but rather directly pass to the list of $SC(g,\rho )$ resulted
from the above proposition. It is based on the list of little conformal Lie
superalgebras given in Refs.[8], [9] (which in turn is based on Kac's
classification of simple Lie superalgebras). There are three classical series,
one continuous family and two exceptional cases related with octonions.

$\bullet $ Case $I:$ $osp(N|2;{\bf R})$, $N = 1,2,...$, $g = so(N)$, $\rho  =
N.$

For $N = 1$ the corresponding OPE defines the Neveu-Schwarz-Ramond superalgebra
[10], [11], for $N = 2$ it is the superalgebra of Ademollo et al. [12]
involving an {\it so}(2), and for $N\geq 3\ SC(so(N),N)$ has been constructed
by Knizhnik [2] and by Bershadsky [3]. Starting at $N=3$ in order to close the
OPE one has to introduce the quadratic non-linearity :{\it JJ}: in the
right-hand side of $G^i(z)G^j(w)$. In cases $N = 3$ and $N = 4$ it is shown by
Goddard and Schwimmer [4] that the corresponding non-Lie type OPEs can be
obtained by factoring out free fermions and bosons with scale dimensions $1/2$
and 0 from the $N = 3$ and $N = 4$ Lie superalgebras of Refs.[12].

$\bullet $ Case $II:$ $su(1,1|N)$, $N = 3,4,...$, $g = su(N) \oplus  u(1)$,
$\rho  = N \oplus  \bar N$, and

$su(1,1|2)/u(1)$, $g = su(2)$, $\rho  = 2 \oplus  \bar 2.$

While in the later case the superconformal algebra involves an
$\widehat{su}(2)$
Kac-Moody algebra and closes as a Lie superalgebra [12], for all $N \geq  3$
one has to introduce quadratic composites [2], [3]. Then the Jacobi identities
require the central charges for the affine $\widehat{su}(N)$ and $\hat u(1)$ to
be
of opposite signs that generally leads to non-unitarity of the Verma modules.

$\bullet $ Case $III:$ $osp(4^*|2M)$, $M = 1,2,...$, $g = usp(2M) \oplus
su(2)$, $\rho  = (2M,2).$

Meanwhile the case $M = 1$ is isomorphic to the subcase $N = 4$ of the case I,
a whole family of superconformal algebras {\it SC} with $M = 2,3,..$. seems to
be not constructed previously. So we present our results for this case.

Let $\alpha ,\beta ,\gamma ,\delta  = 1,2$ and $A,B,C,D = 1,2,...,2M (M \geq
2)$ be indices in the fundamental representation of {\it usp}(2)
$(\simeq su(2))$ and {\it usp}(2M), respectively. They are raised and lowered
with the help of the symplectic metrics ${\varepsilon}^{\alpha \beta } =
-{\varepsilon}^{\beta \alpha }$, ${\varepsilon}_{\alpha \beta } =
-{\varepsilon}_{\beta \alpha }$, ${\varepsilon}_{\alpha \gamma
}{\varepsilon}^{\beta \gamma } =
\delta ^\beta _\alpha $ and $E^{AB} = -E^{BA}$, $E_{AB} = -E_{BA}$,
$E_{AC}E^{BC} = \delta ^B_A$, respectively. Then a $SC$ is given by the
following supercommutation relations among the Virasoro $L_n$ (associated to
$s\ell(2;{\bf R})$ subalgebra of $osp(4^*|2M))$, $\widehat{su}(2)\oplus
\widehat{usp}(2M)$ Kac-Moody currents
$J_{   m}^{\alpha\beta}=
J_{   m}^{\beta\alpha}$
and
$J_{   m}^{AB}=
J_{   m}^{BA}$
(associated to $g)$ and supercurrents
$G_{   r}^{\alpha A}$ (associated to $\rho$)
\medskip
$$
[L_m,L_n] = (m-n)L_{m+n} + {c\over 12} m(m^2-1)\delta_{m+n,0},
\eqno{(1a)}$$
\medskip
$$
[J_{m }^{\alpha\beta} ,J_{   n}^{\gamma \delta} ] =
{\varepsilon}^{\beta \gamma }J_{   m+n}^{\alpha\delta } +
{\varepsilon}^{\alpha \delta }
J_{   m+n}^{\beta\gamma} +
{\varepsilon}^{\alpha \gamma }
J_{   m+n}^{\beta\delta} + {\varepsilon}^{\beta \delta }
J_{   m+n}^{\alpha\gamma}
$$
$$
- k_2({\varepsilon}^{\alpha \gamma }{\varepsilon}^{\beta \delta } +
{\varepsilon}^{\beta \gamma }{\varepsilon}^{\alpha \delta })m\delta _{m+n,0},
\eqno{(1b)}
$$
\medskip
$$[J_{   m}^{AB},J_{   n}^{CD}] =
E^{BC}J_{   m+n}^{AD} + E^{AD}
J_{   m+n}^{BC} + E^{AC}
J_{   m+n}^{DB} +
E^{BD}J_{   m+n}^{CA}
$$
$$
-k_1(E^{AC}E^{BD} + E^{BC}E^{AD})m\delta_{m+n,0},
\eqno{(1c)}
$$
\medskip
$$
[L_m,
J_n^{\alpha\beta} ] =
-nJ_{m+n}^{\alpha\beta},\qquad
[L_m,J_n^{AB}] =
-nJ_{m+n}^{AB},
\eqno{(1d)}
$$
\medskip
$$[L_m,G_r^{\alpha A}] = ({m\over 2} - r)
G_{m+r}^{\alpha A},
\eqno{(1e)}
$$
\medskip
$$[J_m^{\alpha\beta} ,
G_r^{\gamma A}] = {\varepsilon}^{\beta \gamma }
G_{m+r}^{\alpha A} +
{\varepsilon}^{\alpha \gamma }
G_{m+r}^{\beta A}
\eqno{(1f)},$$
\medskip
$$
[J_m^{AB},
G_r^{\alpha C}] = E^{BC}
G_{m+r}^{\alpha A} + E^{AC}
G_{m+r}^{\alpha B},
\eqno{(1g)}
$$
\medskip
$$\{G_r^{\alpha A},
G_s^{\beta B}\} = 2{\varepsilon}^{\alpha \beta }E^{AB}L_{r+s} +
(r-s)[\sigma _1{\varepsilon}^{\alpha \beta }
J_{r+s}^{AB} + \sigma_2E^{AB}
J_{r+s}^{\alpha\beta}]
$$
$$+ b(r^2 - {1\over 4}){\varepsilon}^{\alpha \beta }E^{AB}\delta_{r+s,0} +
\gamma (JJ)_{r+s}^{\alpha A,\beta B},
\eqno{(1h)}
$$

\noindent
where the normal ordered quadratic term reads as follows
\medskip
$$
(JJ)_m^{\alpha A,\beta B} =
2{\varepsilon}^{\alpha \beta }
E^{AB}\{{\varepsilon}_{\alpha _1\beta _1}{\varepsilon}_{
\alpha _2\beta _2}(J^{\alpha _1\alpha _2}J^{\beta _1\beta _2})_m
- E_{A_1B_1}E_{A_2B_2}(J^{A_1A_2}J^{B_1B_2})_m\}
$$
$$
+8(J^{\alpha \beta }J^{AB})_m
+ 4{\varepsilon}^{\alpha \beta }E_{CD}\{(J^{AC}J^{BD})_m - (J^{BC}J^{AD})_m\},
\eqno{(2a)}
$$
\noindent
where
\medskip
$$
(J^{AB}J^{CD})_m = \sum _{n\in {\bf Z}}:
J_n^{AB}J_{m-n}^{CD}:
\eqno{(2b)}
$$

\noindent
and similar for $(J^{\alpha \beta }J^{\gamma \delta })_m$. The central charges
$k_1$, $k_2$, $c$ and $b$ and the structure constants $\sigma _1$, $\sigma _2$
and $\gamma $ are fixed by the Jacobi identities to be

$$k_1 = - (k + 2M+4)/2,
k_2 = k,
\eqno{(3a)}
$$

$$b = - {k(k+2M+4)\over k-2M+4},
\eqno{(3b)}
$$

$$
c = - {3k(k+2M+4)\over k-2M+4} + {6k+(2M+1)(M-2)(k+2M+4)\over k-2M+4},
\eqno{(3c)}
$$

$$\sigma _1 = - {2k\over k-2M+4},\qquad
\sigma _2 = {k+2M+4\over k-2M+4},
\eqno{(3d)}
$$

$$\gamma  = {1\over 2(k-2M+4)}.
\eqno{(3e)}
$$

\noindent
Here $k$ is the only free parameter which determines the values of all the
central charges. One sees $k_1$ and $k_2$ have opposite signs that leads to
non-unitarity similar to the case II (for $N\geq 3)$. To check up on the Jacobi
a crucial part is played by the Fierz-type identity
\medskip
\noindent
$${\varepsilon}^{\alpha \beta }\delta ^\gamma _\delta  +
{\varepsilon}^{\beta \gamma }\delta ^\alpha _\delta  +
{\varepsilon}^{\gamma \alpha }\delta ^\beta _\delta  = 0
\eqno{(4)}
$$
\noindent
which follows from the fact that the antisymmetrization in three two-valued
indices gives zero.

$\bullet $ Case IV: $D^1(2,1;\alpha )$, $g = su(2) \oplus  su(2)$, $\rho  =
(2,2)$ (one-parameter family; $\alpha  \rightarrow  1/\alpha $ gives an
isomorphism).

For $M = 1$ the same identity (4) holds also for capital latin indices {\it A},
$B$, $C$, $D$ and there appears a possibility to have got one more free
parameter, say $\alpha $, in the algebra (1). Then two Kac-Moody central
charges $k_1$ and $k_2$ for "left" and "right" affine $\widehat{su}(2)$ become
independent (in fact $\alpha $ governs a "left-right asymmetry") and it makes
it possible to have both $k_1$ and $k_2$ simultaneously positive integer. The
resulting non-linear superconformal algebra is obtained in Ref.[4] by factoring
out free fermions and bosons from the Lie superalgebra of Refs.[13]. The
unitary highest weight representations are studied and applied to superstring
compactification in Ref.[5].

$\bullet $ Case $V:$ $G(3)$, $g = G_2$, $\rho  = 7.$

Consider an octonionic algebra ${\bf O}$ with seven imaginary units $e^a$, $a =
1,2,...,7$, and a composition law
\medskip
\noindent
$$e^ae^b = -\delta ^{ab} + \Gamma ^{abc}e^c,
\eqno{(5)}
$$

\noindent
where $\Gamma ^{abc}$ is a totally skew $G_2$-invariant tensor. Thanks to the
alternativity of ${\bf O}$, $\Gamma $ obeys a crucial identity
\medskip
\noindent
$$\Gamma ^{abe}\Gamma^{cde} + \Gamma^{cbe}\Gamma^{ade} = 2\delta ^{ac}\delta
^{bd} -
\delta ^{ad}\delta ^{bc} - \delta ^{ab}\delta ^{cd}.
\eqno{(6)}
$$

An exceptional $N=7$ superconformal algebra $SC(G_2,7)$ involves the Virasoro
$L_m$, $\hat G_2$ Kac-Moody currents
$J_m^{ab} = -J_m^{ba}$ satisfying the property (actually it defines the
projection onto $G_2
\subset  so(7)$; for the properties of $G_2$ see $e.g$. [14], or [15]-[17]
where the supergroups $G(3)$ and $F(4)$ are described in detail)
\medskip
\noindent
$$\Gamma ^{abc}J_m^{bc} = 0,
\eqno{(7)}
$$
\noindent
and supercurrents $G_r^a$
in the fundamental seven-dimensional representation of $G_2$. Then our result
for the superalgebra reads as follows (we omit the obvious commutation
relations with $L_m)$
\medskip
\noindent
$$[J_m^{ab},
J_n^{cd}] = \delta ^{bc}
J_{m+n}^{ad}+ \delta^{ad}
J_{m+n}^{bc}- \delta^{ac}
J_{m+n}^{bd}
- \delta^{bd}J_{m+n}^{ac}
$$
$$
+ {1\over 3} \Gamma ^{abe}\Gamma^{cdf}
J_{m+n}^{ef}- k(\delta ^{ac}\delta ^{bd} - \delta ^{ad}\delta ^{bc} -
{1\over 3}\Gamma ^{abe}\Gamma^{cde})m\delta_{m+n,_0}
\eqno{ (8a)}
$$
\medskip
\noindent
$$[J_m^{ab},G_r^c] = \delta ^{bc}G_{m+r}^a
- \delta ^{ac}G_{m+r}^b
+ {1\over 3} \Gamma^{abe}\Gamma^{cde}G_{m+r}^d,
\eqno{(8b)}
$$

$$\{G_r^a,G_s^b\} = 2\delta ^{ab}L_{r+s} + \sigma (r-s)
J_{r+s}^{ab}
+ b(r^2 - {1\over 4})\delta ^{ab}\delta_{r+s,0}
+ \gamma  P_{cd,ef}^{ab}
(J^{cd}J^{ef})_{r+s},
\eqno{(8c)}
$$
\noindent
where the tensor $P$ is uniquely determined from the Jacobi identities
\medskip
\noindent
$$P_{ab,cd} = \{\lambda _{ab},\lambda _{cd}\} + {4\over 3}
(\delta_{ac}\delta _{bd} - \delta _{ad}\delta _{bc} - {1\over 3}
\Gamma_{abe}\Gamma_{cde}) {\bf 1},
\eqno{(9)}
$$

\noindent
and $\lambda _{ab}$ are $G_2$ matrices in the fundamental representation

$$\lambda _{ab,cd} = \delta _{ac}\delta _{bd} - \delta _{ad}\delta _{bc} -
{1\over 3} \Gamma_{abe}\Gamma_{cde}.
\eqno{(10)}
$$

\noindent
All the constant parameters $c,b,\sigma $ and $\gamma $ are uniquely fixed by
associativity to be the following rational functions of the $\hat G_2$
Kac-Moody level $k:$
\medskip
\noindent
$$c = {9k\over 2} + {2k\over k+3},
\eqno{(11a)}
$$
\medskip
\noindent
$$b = k\sigma  = {3k\over 2}(1 - {4\over 3(k+3)}),
\eqno{(11b)}
$$
\medskip
\noindent
$$\sigma  = {3\over 2}(1 - {4\over 3(k+3)}),
\eqno{(11c)}
$$
\medskip
\noindent
$$\gamma  = {3\over 16(k+3)}.
\eqno{(11d)}
$$

\noindent
This time a crucial role in checking up the Jacobi is played by the octonionic
identity (6).

There exists a natural embedding of the $N=1 (NSR)$ superconformal algebra in
our exceptional $N=7$ algebra (more detailed discussion for the $N=8$
exceptional case see in Ref.[1]). Choose the supergenerator $G_r$ of the $N=1$
superalgebra we are about to construct to be equal to
$G_r^1$ and redefine the Virasoro generators according to
\medskip
\noindent
$$\tilde L_m = L_m + {\gamma \over 2}
P_{ab,cd}^{11}(J^{ab}J^{cd})_m.
\eqno{(12)}
$$

\noindent
Then it is easy to see that $\tilde L_m$ and $G_r$ form an $N=1$ superalgebra
with the Virasoro central charge
\medskip
\noindent
$$\tilde c = {9k\over 2}(1 - {4\over 3(k+3)}).
\eqno{(13)}
$$

\noindent
The structure of new Virasoro generators (12) is reminiscent of the GKO coset
construction for $G_2/su(3).$

Note that central charges $c,\tilde c$ and $b$ are positive when the Kac-Moody
level $k$ is positive integer, the obvious necessary condition for unitarity
thus being satisfied. The same also holds for the next case VI.

$\bullet $ Case $VI:$ $F(4)$, $g = Spin(7)$, $\rho  = 8_s.$

The exceptional $N=8$ conformal superalgebra {\it SC} (Spin$(7),8_s)$
associated with $F(4)$ has been constructed in our previous paper [1].

To summarize, in our classification of superconformal OPEs in two dimensions
there are three classical families, one continuous family and two exceptional
cases with $N=7$ and $N=8$. Of special interest are the problems of
constructing representations and field-theoretic realizations of the
exceptional algebras. Probably they could be obtained via the quantum
Drinfeld-Sokolov reduction based on the underlying little superalgebras
$G(3)$ and $F(4).$

\vfill\eject

\centerline{{\bf Note Added}}

In this paper we have considered superconformal algebras with the standard
spin-statistics relation. Recently so-called quasisuperconformal algebras have
appeared in the context of the quantum Drinfeld-Sokolov reduction and
related approaches [16]-[18]. They involve bosonic currents of dimension $3/2$
in a representation $\rho $ of some Lie algebra $g$ along with the
energy-momentum and $\hat{g}$ $KM$ currents. A list of possible pairs
 $(g,\rho )$
may be completed on the basis of Cartan's classification of symmetric
subalgebras of simple Lie algebras. One requires the symmetric subalgebra to be
isomorphic to $su(2) \oplus  g$ and the supplementary subspace to transform in
its representation $(2,\rho )$ with $\rho ^{\bf C}$ possessing a
$g^{\bf C}$-invariant symplectic metric. As a result one finds eight relevant
cases $(g,\rho $ and underlying simple Lie algebras are presented): Case I:
$u(N)$, $N \oplus  \bar N$, $su(N+2)$; Case II: {\it sp}(2M),  2M,
$sp(2M+2)$; Case III: $su(2) \oplus  so(N)$, $(2,N)$, $so(N+4)$; Case IV:
{\it su}(2), $dim\rho  = 4$, $G_2$; Case $V: sp(6)$, $dim\rho  = 14$, $F_4$;
Case VI:$su(6)$,$\rho=20$,$E_6$;Case VII:$so(12)$,$\rho=32^\prime$, $E_7$;
Case VIII
:$E_7$,$\rho=56$,$E_8$.Quasisuperconformal algebras corresponding to
 the Cases I and II have been
studied in Refs.[16] and [17]. The Case III represents a bosonic counterpart of
our Case III superconformal algebras based on $osp(4^*|2M)$. Five exceptional
cases will be concidered in a separate publication . However it should be
stressed that there are difficulties in the Cases $II-VIII$ with imposing the
hermiticity condition, since representations $\rho$ are pseudoreal and,in
particular, it is not possible to define real Euclidean{\it su}(2) spinors.

Then two series of ${\bf Z}_2\times {\bf Z}_2$-graded superconformal algebras
have been introduced in Ref.[18]. They containe affine superalgebras
$\widehat{su}(m|n)$ and $\widehat{osp}(N|2M)$ and in fact combine
 superconformal and
quasisuperconformal algebras together. Our results suggest the existence of a
third series of this type combining the Cases III from our two lists.
It containes affine $\widehat{su}(2)\oplus \widehat{osp}(N|2M)$.
\vskip 2truecm
\noindent \bf Acknowledgement\rm \ \
We are grateful to W.Troost for sending us a copy of the preprint [18]. \ \
E.S. Fradkin thanks Prof. J. Fr\"ohlich and the ``Zentrum f\"ur Theoretische
Studien'' at ETH for hospitality and also the SNF (Swiss National Fonds) for
financial support.

\vfill\eject
\centerline{{\bf References}}

\noindent
1. E.S.Fradkin and V.Ya.Linetsky, $ICTP/91/348$ (Phys. Lett. $B$, to be
published).

\noindent
2. V.G.Knizhnik, Theor. Math. Phys. 66 (1986) 68.

\noindent
3. M.Bershadsky, Phys. Lett. B174 (1986) 285.

\noindent
4. P.Goddard and A.Schwimmer, Phys.Lett. B214 (1988) 209.

\noindent
5. $M.G\ddot u$naydin, J.L.Petersen, A.Taormina and A.Van Proeyen, Nucl. Phys.
B322 (1989)  402.

\noindent
6. P.Ramond and J.H.Schwarz, Phys. Lett. B65 (1976) 75.

\noindent
7. E.S.Fradkin and V.Ya.Linetsky, Classification of superconformal algebras in
two dimensions, in preparation.

\noindent
8. W.Nahm, Nucl. Phys. B135 (1978) 149.

\noindent
9. $M.G\ddot u$naydin, G.Sierra and P.K.Townsend, Nucl. Phys. B274 (1986) 429.

\noindent
10. P.Ramond, Phys. Rev. D3 (1971) 2415.

\noindent
11. A.Neveu and J.H.Schwarz, Nucl. Phys. B31 (1971) 86.

\noindent
12. M.Ademollo et al., Phys. Lett. B62 (1976) 105; Nucl. Phys. B111 (1976) 77;
B114 (1976) 297.

\noindent
13. A.Sevrin, W.Troost and A.Van Proeyen, Phys. Lett. B208 (1988) 447;
K.Schoutens, Nucl. Phys. B295 (1988) 634.

\noindent
14. R.E.Behrends, J.Dreitlein, C.Fronsdal and W.Lee, Rev. Mod. Phys. 34 (1962)
5.

\noindent
15. M.Scheunert, W.Nahm and V.Rittenberg, J. Math. Phys. 17 (1976) 1640; B.S.
De Witt and P. van Nieuwenhuizen, J. Math. Phys. 23 (1982) 1953; A.Sudbery, J.
Math. Phys. 24 (1983) 1986.

\noindent
16. A.M.Polyakov, in "Physics and Mathematics of Strings", World Scientific,
1990; M.Bershadsky, Commun. Math. Phys. 139 (1991) 71; F.A.Bais, T.Tjin and P.
van Driel, Nucl. Phys. B357 (1991) 632;
A.Bilal, CERN$-TH.6294/91.$

\noindent
17. L.J.Romans, Nucl. Phys. B357 (1991) 549; F.Defever, Ph.D.Thesis, Leuven
1991.
\noindent

18. F.Defever, W.Troost and Z.Hasiewicz, Preprint KUL-TF-91-26\&ITP-UWr91-779
(Phys. Lett. B273 (1991) 51).
\end{document}